\documentclass[aps,prl,twocolumn,groupedaddress,showpacs,floatfix,superscriptaddress,longbibliography]{revtex4-2}
\usepackage[plainpages=false,pdfpagelabels,colorlinks=true,linkcolor=red,urlcolor=blue,citecolor=blue,pdftitle={Title},pdfauthor={},pdfdisplaydoctitle=true,pdfduplex=DuplexFlipLongEdge]{hyperref}
\usepackage[a4paper,margin=1in,footskip=0.25in]{geometry}
\usepackage{epsfig}
\usepackage{graphicx}
\usepackage[utf8]{inputenc}
\usepackage{amsmath,amssymb}
\usepackage{physics}
\usepackage[dvipsnames]{xcolor}

\usepackage[normalem]{ulem}

\begin{document}
\title{Supersolid phase in the diluted Holstein model}
\author{Jingyao Meng}
\affiliation{School of Physics and Astronomy, Beijing Normal University, Beijing 100875, China\\}
\author{Yuxi Zhang}
\affiliation{School of Physics and Astronomy, University of Minnesota, Minneapolis, Minnesota 55455, USA}
\author{Rafael M. Fernandes}
\affiliation{School of Physics and Astronomy, University of Minnesota, Minneapolis, Minnesota 55455, USA}
\author{Tianxing Ma}
\email{txma@bnu.edu.cn}
\affiliation{School of Physics and Astronomy, Beijing Normal University, Beijing 100875, China\\}
\affiliation{Key Laboratory of Multiscale Spin Physics (Ministry of Education), Beijing Normal University, Beijing 100875, China\\}

\author{R. T. Scalettar}
\affiliation{Department of Physics and Astronomy, University of California,
Davis, California 95616,USA}

\begin{abstract}
The Holstein model on a square lattice at half-filling has a well-established finite temperature
phase transition to an insulating state with long range charge density wave (CDW) order.
Because this CDW formation suppresses pairing, a superconducting (SC) phase emerges only with doping.
In this work, we study the effects of dilution of the local phonon degrees of freedom in the Holstein model while keeping the system at half filling.
We find not only that the CDW remains present up to a dilution fraction $f \sim 0.15$, but also
that long range pairing is  stabilized with increasing $f$, resulting in a supersolid regime centered at $f \approx 0.10$, where
long range diagonal and off-diagonal correlations
coexist.
Further dilution results in a purely SC phase, and ultimately in a normal metal.
Our results provide a new route to the supersolid phase via the introduction
of impurities at fixed positions which both increase quantum fluctuations
and also are immune to the competing tendency to phase separation often observed in the doped case.
\end{abstract}

\date{Version 16.0 -- \today}

\maketitle

\emph{Introduction.}~
A regular arrangement of particle positions on the one hand, and particle mobility on
the other, are typically competing tendencies.
Nevertheless,
the possibility of the coexistence of crystalline order and the most
extreme type of transport, superfluid flow, was considered in Helium
almost seven decades ago \cite{penrose1956bose,andreev1969quantum,chester1970speculations,leggett1970can}.
Although experimental confirmation in Helium was controversial \cite{kim2004probable,xiao2012anomalous,kim2012absence},
studies of several model Hamiltonians exhibiting such ``supersolid'' (SS) properties have
been reported \cite{scalettar1995simultaneous,van1995quantum,frey1997critical,wessel2005supersolid,melko2005supersolid,laflorencie2007quantum}.
In electronic models, a supersolid phase is often identified by the coexistence between an insulating phase that breaks translational symmetry (such as a charge density wave, CDW) and a phase displaying off-diagonal long-range order (such as superconductivity, SC). 
Until now, supersolidity continues to be an intriguing area of quantum matter research, including novel solid state realizations such as ``spin'' supersolids
like  Na$_2$BaCo(PO$_4$)$_2$ \cite{xiang2024giant} and
K$_2$Co(SeO$_3$)$_2$ \cite{chen2024phase}. 
More broadly, continuous systems such as ultracold quantum gases offer the new platforms to achieve the realization of supersolidity \cite{li2017stripe,leonard2017supersolid,tanzi2019supersolid,norcia2021two}, and this kind of supersolidity with continuous translational symmetry breaking is distinguished from the one in the lattice Hamiltonian with discrete translational symmetry.

The Holstein Hamiltonian \cite{Holstein59}, where the electronic site density
is coupled to a local (dispersionless) oscillator mode, offers a possible arena
in which to explore supersolid behavior. At half-filling,
$\langle n \rangle =1$, the low temperature phase exhibits CDW order
and is insulating. The transition temperature  $T_{\rm cdw}$ has been determined on a square lattice
both via quantum Monte Carlo (QMC) simulations \cite{hohenadler16,Zhang19,xing21}
and machine learning methods \cite{costa2017principal}, thus firmly establishing the existence of a
finite temperature transition to a commensurate solid.

One of the key properties of the Holstein Hamiltonian in the dilute limit is polaron formation,
in which an electron localized to a site rearranges the local phonon configuration.
As a result, when the electron moves, it must carry the lattice distortion along.
Since Holstein polarons are quite heavy, as their density
increases to the point where condensation into a superconducting
phase becomes possible, the resulting critical temperatures
are very low \cite{bradley2021superconductivity}.
While a SC phase has been established \cite{nosarzewski2021superconductivity,ly2023comparative,araujo2022two},
no coexistence of SC and CDW order has been
observed in QMC studies of the Holstein model as the filling is
varied \cite{vekic1992charge,weber2018two,PhysRevLett.120.187003},
although a projector renormalization group method does report a supersolid
at very weak coupling \cite{sykora2009coexistence}.

Supersolids are linked to the high mobility of quantum fluctuation vacancies in the
otherwise ordered background of particle positions. A rather delicate balance is thus required:
the doping must be light
enough to allow the rigid pattern to coexist with the holes, yet sufficient
quantum fluctuations are needed to form a condensate. Moreover, the tendency of the vacancies to phase-separate from the ordered background must also be avoided. Here we demonstrate, using QMC simulations, a new route to supersolid behavior in an electron-phonon model
in which the number of fermions remains commensurate (i.e., no doping is introduced), but the dilution
is instead introduced in the bosonic degrees of freedom to which they
couple. Through this way, it provides an interesting possibility to establish the physical system for supersolids both experimentally and theoretically. 
We show that
measurements of the charge structure factor, superfluid susceptibility,
compressibility, and spectral function form a consistent picture of
the traversal of a CDW-SS-SC-normal sequence of phases
as the dilution of the phonon modes increases.
Our approach bears conceptual similarities to the investigations of
supersolidity in Helium
in that the dilution of phonon modes provides a random landscape
analogous to the nanometer scale pore size of the Vycor glass in which the
solid Helium is placed.

\emph{Model and method.}~
The Holstein Hamiltonian \cite{Holstein59},
\begin{align} 
\label{eq:Holst_hamil}
\nonumber \cal{\hat H} = & -t \sum_{\langle \mathbf{i}, \mathbf{j}
  \rangle, \sigma}
  \big(\hat d^{\dagger}_{\mathbf{i} \sigma} \hat d^{\phantom{\dagger}}_{\mathbf{j} \sigma} + {\rm H.c.} \big)
- \mu \, \sum_{\mathbf{i}, \sigma}
\hat n^{\phantom{\dagger}}_{\mathbf{i}, \sigma} \\ & + \omega_{0} \sum_{
  \mathbf{i} } \hat a^{\dagger}_{\mathbf{i}}
\hat a^{\phantom{\dagger}}_{\mathbf{i}} + \sum_{\mathbf{i}, \sigma}
g^{\phantom{\dagger}}_{\mathbf{i}} \, \hat n^{\phantom{\dagger}}_{\mathbf{i}\sigma} \big(
\hat a^{\dagger}_{\mathbf{i}} + \hat a^{\phantom{\dagger}}_{\mathbf{i}} \big) \,
,
\end{align}
describes electrons of spin $\sigma = \uparrow,\downarrow$ hopping between
nearest-neighbor sites $\langle {\bf i,j} \rangle$
and interacting with a local phonon mode on each site.
In Eq.~(\ref{eq:Holst_hamil}),
$\hat d^{\dagger}_{\mathbf{i} \sigma}(\hat d^{\phantom{\dagger}}_{\mathbf{i}\sigma})$
are fermion creation (destruction) operators with the given site and spin indices and
with corresponding number operator $\hat n_{{\bf i}\sigma}^{\phantom{\dagger}}
= \hat d^{\dagger}_{\mathbf{i} \sigma}\hat d^{\phantom{\dagger}}_{\mathbf{i}\sigma}$, whereas
$\hat a^{\dagger}_{\mathbf{i}} (\hat a^{\phantom{\dagger}}_{\mathbf{i}})$
are phonon creation (destruction) operators. The parameters $t, \omega_0$
are the hopping energy (which we set to be our energy scale,  $t=1$ ) and phonon frequency,  respectively,
$g_{\bf i}$ is the electron-phonon coupling, and
$\mu$ is the chemical potential.
For the clean system with $g_{\bf i}=g$, the chemical potential corresponding to half-filling is $\mu_{0}=-2g^{2}/\omega_{0}$.

We introduce dilution by allowing for a random, site-dependent, electron-phonon coupling  $g_{\bf i}$,
such that the coupling vanishes on a fraction $f$ of the sites:
\begin{align}
g_{\bf i}=
\left\{
\begin{array}{rc}
 g  & \qquad 1-f  \\
 0  & \qquad  f.
\end{array}
\right.
\end{align}
We consider $N=L \times L$ square lattices.
Simulations are typically averaged over five to ten realizations of the random locations.
If $f N$ is not an integer, we calculate a weighted average of its adjacent integers,
providing further disorder averaging.
Further discussion is found in the Supplemental Material \cite{SP} (see also Refs. \cite{berche2004bond,PhysRevB.58.2740,PhysRevE.58.2938,PhysRevLett.81.252,Machida1984,PhysRevB.82.014521,kaufmann23,PhysRevLett.126.056402,PhysRevB.101.085111,HUANG2023108863,PhysRevLett.130.226001,PhysRevA.74.013608,PhysRevB.76.144513} therein).

We investigate the competition between SC and CDW order through the determinant quantum Monte Carlo (DQMC) method \cite{Blankenbecler81,PhysRevB.40.197,PhysRevLett.120.187003},
an unbiased auxiliary-field approach for the computation of finite-temperature properties. We perform the usual mapping of the quantum oscillator degrees of freedom onto a path integral in imaginary time by discretizing the inverse temperature $\beta=\Delta\tau L_{\tau}$ \cite{Creutz1981ASA}, then the degrees of freedom of the fermions moving in this fluctuating space and imaginary time phonon field can be integrated out analytically.
Since the fermionic operators appear quadratically in the Holstein Hamiltonian,
they can be integrated out analytically. This results in an action that is the square of the determinant
of a matrix and that depends on the space and imaginary-time dependent quantum phonon field, which is
then sampled stochastically.
The square in the determinant arises because the determinants of the up and down fermions are identical, i.e., there is no sign problem. 
The discretization mesh $\Delta\tau$ of the inverse temperature $\beta=1/T$ was chosen small enough so that the ``Trotter errors'' are smaller than those associated with the statistical sampling.  We set the system
to half filling with charge density $\langle n \rangle=1$, and define the dimensionless electron-phonon coupling 
$\lambda_{D}=g^{2}/(zt\omega_{0})$, where $z=4$ is the coordination number for the square lattice. 
In this work, we mainly focus on systems with $g=1$, $\omega_{0}=1$  ($\lambda_D=0.25$) and $g=2$, $\omega_{0}=3$ ($\lambda_D=0.33$), 
but also provide some results for additional $\omega_{0}$ values.

In order to discern CDW and SC phases,
we define the equal-time, real-space charge correlation function,
\begin{equation}
c({\bf r}) = \langle(\hat n_{{\bf i}\uparrow}+\hat n_{{\bf i}\downarrow})(\hat n_{{\bf i+r}\uparrow}+\hat n_{{\bf i+r}\downarrow})\rangle,
\label{eq:cdw_real}
\end{equation}
and its Fourier transform $S(\bf{q})$,
\begin{equation}
S({\bf q})=\sum_{{\bf r}}e^{i{\bf q}\cdot({\bf r})} c({\bf r}).
\label{eq:cdw_struc}
\end{equation}
At commensurate filling $\langle n \rangle=1$,
the structure factor is sharply peaked at ${\bf q}=(\pi,\pi)$
owing to the perfect nesting of the noninteracting Fermi surface. Hence,  we define
$S_{\rm cdw} \equiv S(\pi,\pi)$.

The $s$-wave pairing susceptibility is 
\begin{equation}
P_{s}=\frac{1}{N}\int_{0}^{\beta}d\tau\langle\hat \Delta(\tau)\hat \Delta^{\dagger}(0)\rangle,
\label{eq:sc_susc}
\end{equation}
where $\hat \Delta(\tau)=\sum_{\bf i}\hat c_{{\bf i}\downarrow}(\tau)\hat c_{{\bf i}\uparrow}(\tau)$.
We study SC via the (imaginary time integrated) susceptibility, rather than the equal time structure factor, because the former provides a more sensitive measure for pairing order, which is less robust than charge order.

$S({\bf q})$ and $P_s$ are both normalized in such a way that
in a high-temperature or otherwise disordered phase, their values are independent of
lattice size $N=L^2$ as long as $\xi \lesssim L$, where $\xi$ is the correlation length.
However, in an ordered phase with
$\xi \gtrsim L$,  $S_{\rm cdw}$ and $P_s$ grow linearly with $N$.
This provides an immediate, albeit somewhat crude, means by which
long range order can be discerned.

\begin{figure}[t]
\includegraphics[height=1.60in,width=3.00in]{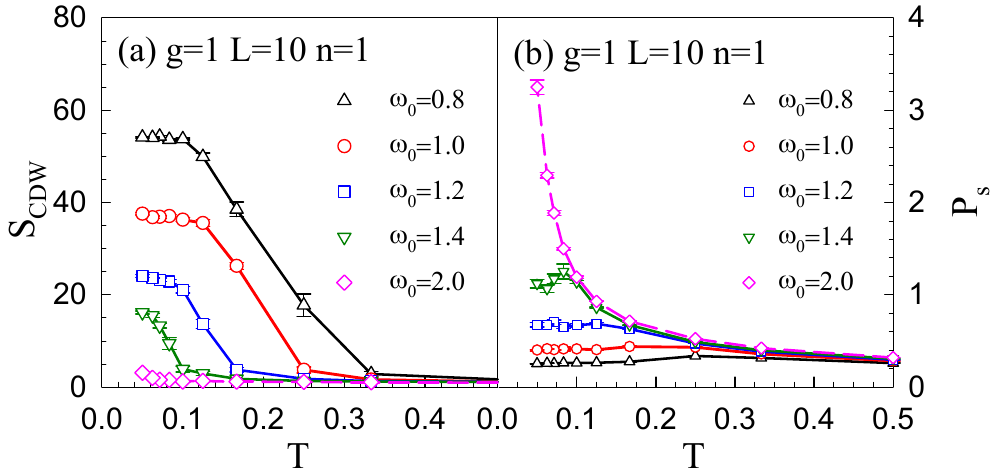}
\caption{DQMC results in the absence of dilution ($f=0$).
(a) CDW structure factor $S_{\rm cdw}$ on an $N=10\times10$ lattice as a function of temperature $T$
for $g=1$ and half-filling, $\langle n \rangle=1$.
For smaller $\omega_0$ there is a sharp increase in $S_{\rm cdw}$ for $T<T_{\rm{cdw}}$,
providing preliminary evidence of an ordered CDW phase.
However, as $\omega_0$ increases, charge correlations are diminished.
(b) $s$-wave pairing susceptibility $P_{s}$ as a function of $T$ for the same parameters.
$P_{s}$ increases with increasing $\omega_0$.
For the largest phonon frequency, where $S_{\rm cdw}$ essentially vanishes, the pairing
susceptibility grows dramatically at low $T$.
For fixed $g/t=1$, the phonon frequencies $\omega_0/t = 0.8, 1.0, 1.2, 1.4, 2.0$ correspond
to dimensionless couplings $\lambda_D = g^2/(z t\omega_0) = 0.313, 0.250, 0.208, 0.179, 0.125$
respectively.
}
\label{fig:CDW-SC-D0}
\end{figure}

\emph{Results.}~
We start from the undiluted case, $f=0$.
In Fig.~\ref{fig:CDW-SC-D0} we show
$S_{\rm cdw}$
and $P_{s}$ as functions of temperature $T$ for electron-phonon coupling $g=t$ and 
varying phonon frequency $\omega_0$, as well as linear lattice size $L=10$ ($N=100$). 
For $\omega_0 \lesssim 1.4\,t$ ($\lambda_D \gtrsim 0.18$), $S_{\rm cdw}$ 
rises sharply below a critical temperature that becomes smaller 
as the dimensionless coupling decreases ($\omega_0$ increases).  For small $\omega_0$, 
there is no signal of superconductivity, since $P_s$ is small and 
almost temperature independent.  When the phonon frequency approaches $\omega_0=1.4\,t$, 
the pair structure factor grows as $T$ decreases. This enhancement 
is, however, terminated at the temperature for which $S_{\rm cdw}$ 
rises, reflecting the competition of charge and pairing order. 
For the highest phonon frequency, $\omega_0 = 2\,t$, for which 
$S_{\rm cdw}$ remains small down to at least $T=t/20$, $P_s$ 
shows an especially marked growth at low temperatures. 
That superconductivity is most easily observed in the 
anti adiabatic limit of large $\omega_0$ is an established 
conclusion of prior QMC studies of the Holstein model \cite{bradley2021superconductivity,nosarzewski2021superconductivity}.

Having briefly reviewed the charge and pair correlations in the clean limit
to establish a baseline for comparison, we now explore non zero dilution $f$.
We begin, in Fig.~\ref{fig:CDW-SC}, by choosing two parameter sets:  $g=1$, $\omega_{0}=1$ ($\lambda_D=0.25$); 
and $g=2$, $\omega_{0}=3$ ($\lambda_D=0.33$). In both cases, $\lambda_D$ is greater than the value 
$\lambda_D \sim 0.18$ above which the data of 
Fig.~\ref{fig:CDW-SC-D0} show a sharp rise in $S_{\rm cdw}$ as $T$ is lowered.
Thus, charge correlations are dominant at $f=0$ in these regimes, 
enabling us to investigate 
the effect of dilution on the small pairing correlations 
in a system that begins with a dominant CDW phase in the clean limit. Importantly, since the CDW gap completely gaps out the Fermi surface at half filling, the CDW phase is incompressible (i.e., insulating), characterized by a vanishingly small compressibility; we will return to this point later.

\begin{figure}[t]
\includegraphics[height=2.80in,width=3.00in]{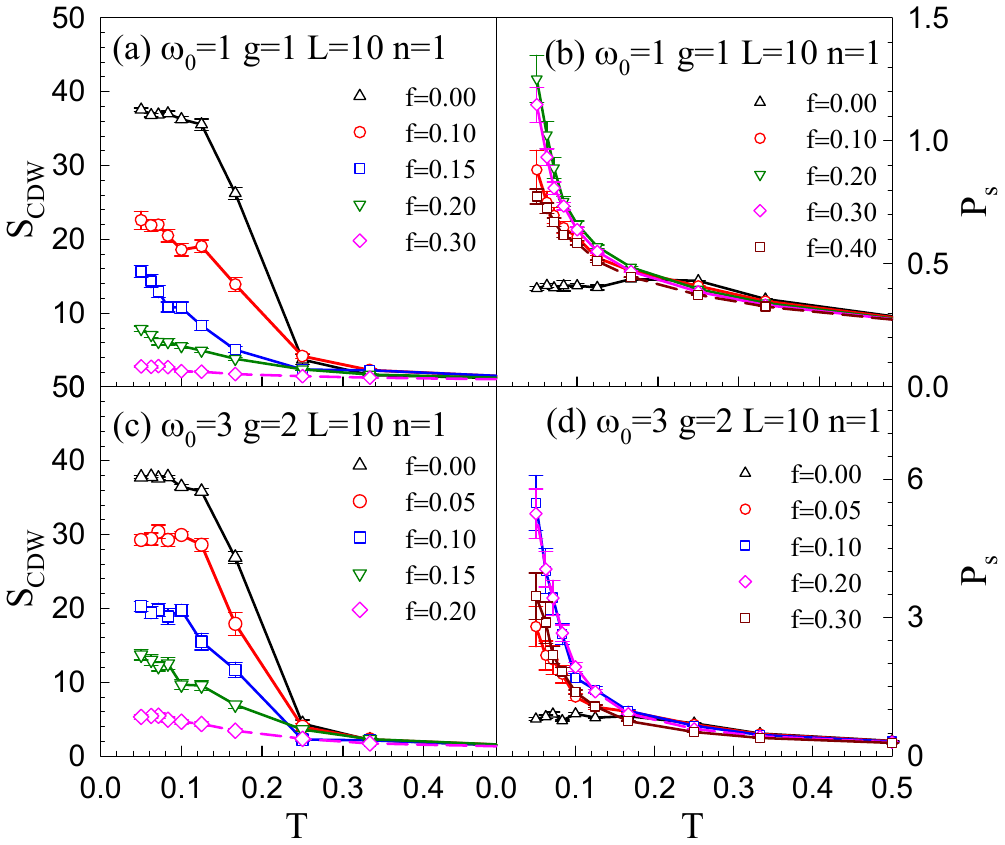}
\caption{CDW structure factor and $s$-wave pairing susceptibility as
functions of $T$ for different dilution fractions $f$.
 Starting from a CDW-dominant situation, $f$ first enhances
and then inhibits $P_{s}$, whereas it always inhibits $S_{\rm cdw}$.
}
\label{fig:CDW-SC}
\end{figure}

As shown in Fig.~\ref{fig:CDW-SC}, dilution
suppresses the CDW phase.  For dilution  $f=0.1$, the value 
of $S_{\rm cdw}$ at $\beta \,t=20$ is reduced by roughly a factor of two.  In addition, the convergence
of $S_{\rm cdw}$ to its ground state value is shifted to
lower temperature. In contrast, the SC behavior is more complex. 
In the clean system, the pairing susceptibility $P_{s}$ slowly decreases with decreasing
temperature at low $T$.  The introduction  of dilution first significantly enhances the 
value of $P_{s}$ and makes it grow sharply as $T\rightarrow 0$.
This effect is strongest at $f \sim 0.1 - 0.2$. 
As $f$ increases further, $P_{s}$ decreases. Therefore,
SC benefits from the destruction of CDW order with a small amount of dilution, 
but eventually too much dilution suppresses the electron-phonon coupling essential for pairing.
While the two parameter sets show the same trends, the pairing correlations for 
larger $\omega_0=3$ [panel (d)] are significantly larger in magnitude 
than those for smaller $\omega_0=1$ [panel (b)].

This nonmonotonicity is reminiscent of the ``superconducting dome'' 
of the cuprates, and of the Hubbard model, where, in a similar way, doping first eliminates long range antiferromagnetic 
order, but too much doping removes the spin-fluctuations providing the pairing ``glue''. 
The analogy is incomplete, since in the Hubbard Hamiltonian there are only electronic degrees of freedom. 
Moreover, the role of remnant CDW fluctuations in mediating the pairing in 
the Holstein model is unclear \cite{bickers1987cdw,PhysRevB.40.197}.

\begin{figure}[t]
\includegraphics[scale=0.38]{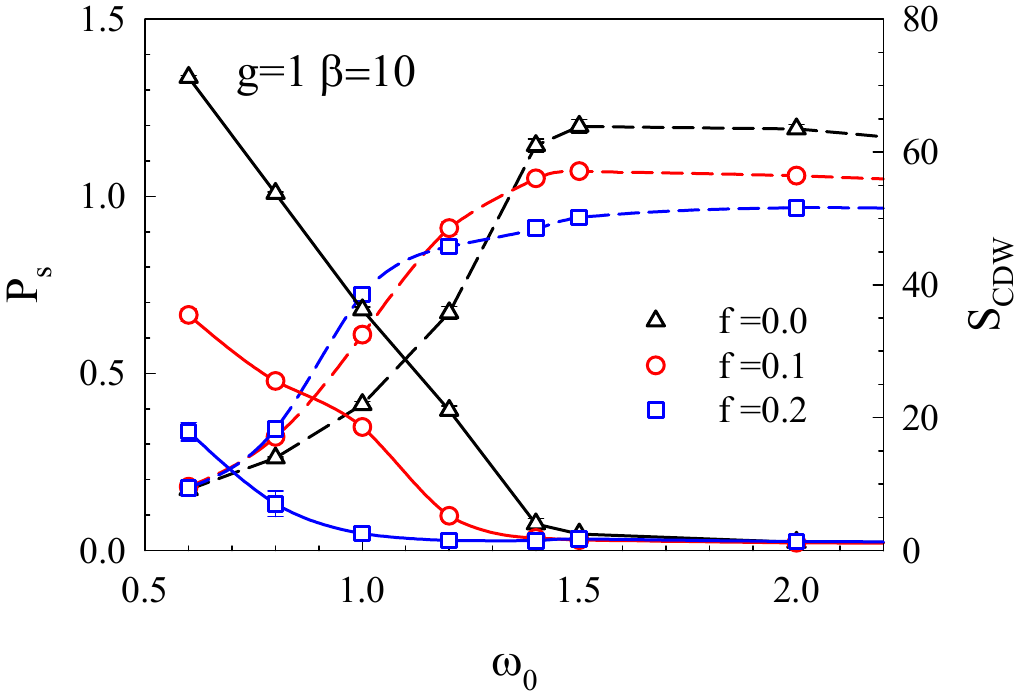}
\caption{
Pairing and CDW correlations as functions of $\omega_{0}$ for different
dilution fractions. Dashed lines represent $P_{s}$ while solid lines represent $S_{\rm cdw}$.
The inverse temperature is fixed at $\beta \, t =10$.
CDW and SC compete, but coexistence is possible in the region of transition
between them.
}
\label{fig:CDW-PS_w}
\end{figure}

An alternative display of the effects of dilution is shown in 
Fig.~\ref{fig:CDW-PS_w}, where $P_{s}$ and $S_{\rm cdw}$ at 
different $f$ are shown as functions of $\omega_{0}$. The 
most evident message, common to all values of $f$, 
is the competition between CDW and pairing order. 
The crossing between the two quantities is shifted to smaller $\omega_0$ with increasing $f$.
Moreover, for small $\omega_0$, 
where CDW order is dominant, $f$ enhances the pairing susceptibility, 
whereas in the absence of CDW order at large $\omega_0$, dilution 
suppresses pairing.

The window of intermediate $\omega_0/t \sim 1$ 
in Fig.~\ref{fig:CDW-PS_w} raises the possibility of a co-existence of 
pairing and CDW orders induced by dilution. 
Such a phenomenon does not occur in the Holstein model
with randomness introduced in the on-site energies \cite{PhysRevB.103.L060501}.
In order to explore whether such a ``supersolid'' phase
exists, we must perform a finite-size scaling study of the two
order parameters.

\begin{figure}[t]
\includegraphics[height=2.80in,width=3.00in]{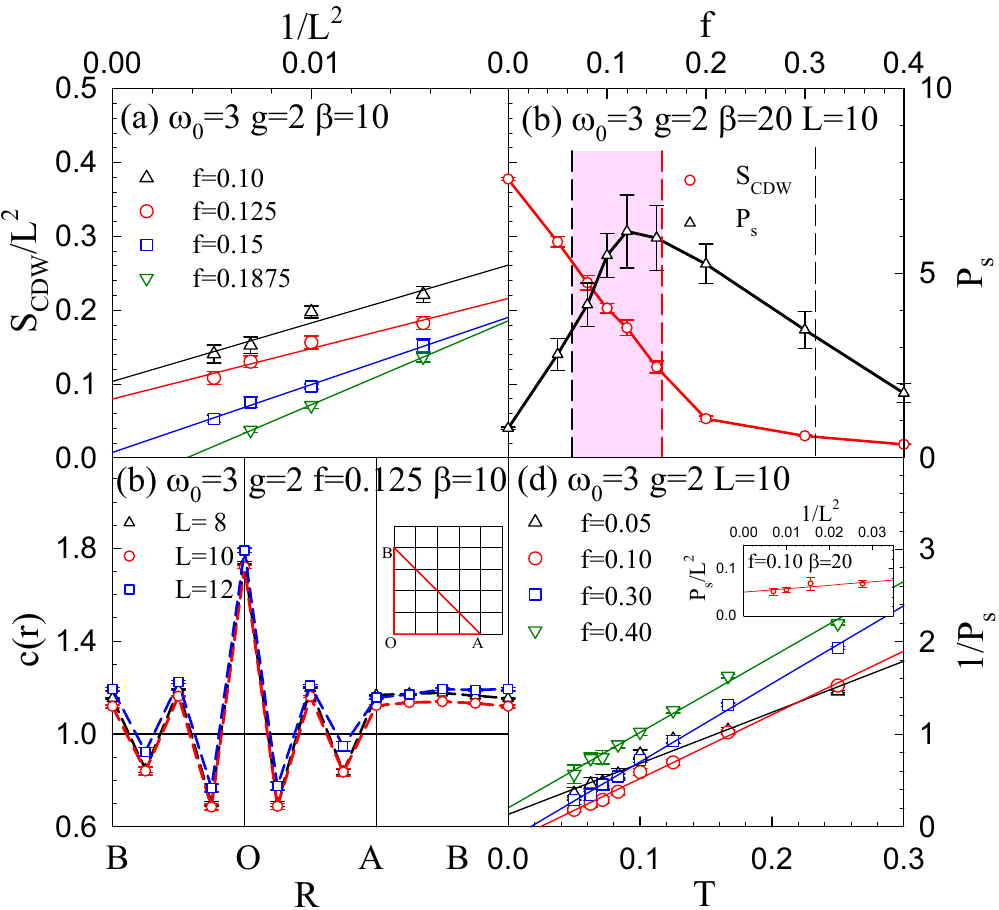}
\caption{(a) Normalized CDW correlations $S_{\rm cdw}/L^{2}$ as a 
function of inverse squared lattice size $L^{-2}$. 
A positive intercept represents long range order.
(b) Pairing and normalized CDW correlations as functions of dilution fraction
at a fixed low temperature ($\beta t=20$). The red dashed line 
demarcates where CDW ends, and the black dashed lines give the range of SC order. 
Thus, the pink shaded region shows the supersolid phase.
(c) Charge correlations $c(r)$ along the path shown in the inset.
(d) $1/P_{s}$ as a function of $T$. For intermediate dilutions $f=0.10$ and $f=0.30$, $1/P_s$ vanishes at finite $T$.
This divergence of the susceptibility signals a SC phase transition.
}
\label{fig:FSS}
\end{figure}

We determine the window of supersolid regime in Fig.~\ref{fig:FSS}. We first 
ascertain the critical dilution that totally destroys CDW order. 
In panel (a), we show the normalized CDW correlations as a function 
of $1/L^{2}$. When increasing $f$ from 2/16 to 3/16, dilution 
suppresses $S_{\rm cdw}/L^{2}$ for each lattice size, and the 
intercept of the curve in the thermodynamic limit, which is the square of the CDW 
order parameter, goes to zero. Indeed, for $f \gtrsim 0.15$, $S_{\rm cdw}/L^{2}$ tends to 
zero  when $L\rightarrow\infty$, suggesting that the CDW correlations are not long ranged and that
there is short range order only. 
The real space charge correlation function $c(r)$ is shown in panel (c).

We now turn to the SC response. 
The reciprocal of the $s$-wave pairing susceptibility, $1/P_{s}$, is shown as 
a function of $T$ in Fig.~\ref{fig:FSS}(d). 
For $f=0.05$, $P_s$ remains finite for all $T$, since $1/P_s$ does not vanish. 
However, for $f=0.10$ and $f=0.30$ there is a nonzero SC critical temperature, signaled by the fact that the extrapolated $1/P_s$ crosses the horizontal axis, that is, the $P_{s}$ curve diverges at $T\rightarrow0$. Increasing dilution further to $f = 0.40$ leads to the vanishing of long range SC order. 
We also show in the insert a finite value of normalized pairing susceptibility 
when extrapolated to $L\rightarrow\infty$ at $f = 0.10$, suggesting the dilution does enhance the SC long-rang order.
We summarize the results in panel (b). 
While a SC phase exists in an intermediate dilution range, $f\in(0.075, 0.35)$, 
CDW order, which is dominant in the clean $f=0$ limit,  remains present up to $f \sim 0.15$.
A supersolid window, where CDW and SC phases coexist, is present and centered around $f \sim 0.10$. 
The evolution of the single particle spectral function provides 
further evidence for the coexistence of CDW and SC order, as we show in the 
Supplemental Material \cite{SP}.

\begin{figure}[t]
\includegraphics[height=2.80in,width=3.00in]{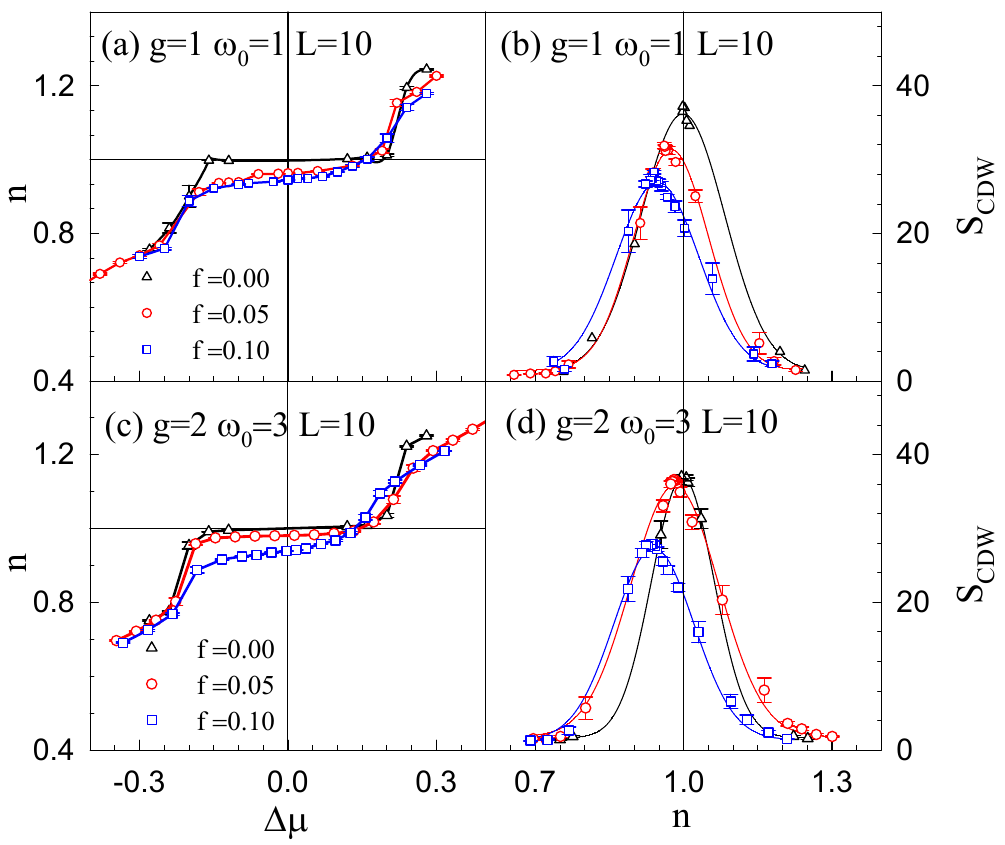}
\caption{Density $ \langle n \rangle$ as a function of the shift in chemical potential $\Delta\mu$
away from $\mu_0 = -2g^2/\omega_0$, which gives half-filling in the clean ($f=0$)
limit. (a) $g=1$, $\omega_{0}=1$ and (c) $g=2$, $\omega_{0}=3$. $S_{\rm cdw}$ as a function 
of density $n$ at (b) $g=1$, $\omega_{0}=1$ and (d) $g=2$, $\omega_{0}=3$. 
The peak of $S_{\rm cdw}$ is shifted from $ \langle n \rangle=1$ by $f$.
 }
\label{fig:gap}
\end{figure}

The ``conventional'' picture of a SS phase is one of the condensation
of mobile vacancies (e.g.,~through a shift of the chemical potential $\mu$)
whose motion does not sufficiently destroy the solid order.
This scenario appears inapplicable here, since although
we dilute the phonon degrees of freedom, we remain at half-filling
for all the data of Fig.~\ref{fig:FSS}.
Figures~\ref{fig:gap}(a) and (c) lend insight into this question by showing the density $\langle n \rangle$ as a 
function of $\Delta\mu= \mu - \mu_0$, for different $f$ values. 
Here, $\mu_{0}=-2g^{2}/\omega_{0}$ is the chemical potential 
for half-filling of the clean system. 
What we observe is that, upon dilution, the plateau in $\langle n \rangle$ no longer occurs at 
commensurate filling $\langle n \rangle=1$.  In other words, the filling $n=1$ is away 
from the filling which gives insulating behavior.  In that sense, our system 
displays ``self-doping''. 
While the data shown are for a single realization,  a discussion of disorder averaging is included in the Supplemental Material \cite{SP}.

Figures~\ref{fig:gap}(b) and (d) provide confirmation of this picture.
$S_{\rm cdw}$  is shown as a function of filling $\langle n\rangle $ for different
dilutions $f$.  The most robust CDW order does not occur at $\langle n \rangle=1$, but rather
at a filling $\langle n \rangle \neq 1$ corresponding to the density at which the compressibility
$\kappa = \partial \langle n  \rangle/ \partial \mu$ vanishes in panels (a) and (c).
Thus the filling $n=1$ can be viewed as being doped away from the filling of largest CDW order. Further insight into this can be obtained by a calculation of the spectral function, shown in the Supplemental Material \cite{SP}.

\emph{Discussion.}~
The experimental search for supersolidity in Helium
 \cite{kim2004probable}, although later rebutted
 \cite{kim2012absence}, had nevertheless an intriguing premise:  by
placing the Helium in the porous environment of Vycor glass, the number
of delocalized vacancies associated with zero point fluctuations of the quantum solid
might be enhanced, thereby increasing the tendency to superfluidity.

Inspired by this general idea, in this work,
using DQMC simulations, we have studied the effect of a fractional dilution $f$
of local oscillators in the Holstein Hamiltonian
on the competition between superconductivity and charge density wave formation.
Through an analysis of the CDW structure factor and SC susceptibility we provide
evidence that a SS phase can be induced even at half filling.  The seemingly surprising occurrence of a SS
without (carrier) dilution is explained by the fact that
that even though the fermion filling $\langle n \rangle$ nominally remains at the
commensurate value $\langle n\rangle=1$ (no ``vacancies''),
the shift in the optimal filling for CDW order away from
half-filling effectively results in vacancies at $\langle n\rangle=1$.

One of the obstacles to conventional supersolid formation is the possibility of
phase separation of the mobile vacancies, a phenomenon known to
obscure supersolid formation for bosonic systems on square lattices \cite{batrouni2000phase}.
The realization of supersolidity presented here avoids that obstacle, because the
phonon dilution pattern is fixed in space.  Thus, the mobile vacancies that are introduced
are necessarily spread out spatially.

It is interesting to ask whether the dilute Holstein model studied here could be realized in materials. While an in-depth analysis is beyond the scope of this work, we note that transition metal dichalcogenides (TMDs) are well known for displaying CDW which, in some cases, coexists with or is in proximity to a SC phase. More specifically, this is the case of two nonmetallic TMDs: $1T$-TiSe$_2$, which displays semiconductinglike transport properties in the CDW state \cite{Morosan2006,Tuti2009}, and $1T$-TaS$_2$, which is believed to be a Mott insulator in the CDW phase \cite{Sipos2008,Nandini2014}. In the case of $1T$-TiSe$_2$, Cu intercalation \cite{Morosan2006} or pressure \cite{Tuti2009} suppresses the CDW phase and promotes SC, with both states overlapping in the phase diagram. Because Cu is introduced between the Ti-Se layers, it is expected to locally affect the crystal structure, which in turn should cause a local change in the electron-phonon interaction, as the atoms will be displaced from their original positions. In the case of  $1T$-TaS$_2$, it has been shown that application of pressure \cite{Sipos2008} leads to a SC phase as the CDW is suppressed, with a possible region of coexistence. To the best of our knowledge, however, SC has not yet been observed in Cu intercalated samples \cite{Nandini2014}.
Recently, it has been confirmed that Pd intercalation induces disorder in the crystal lattice of ErTe3, suppressing CDW formation and leading to a SC ground state \cite{PhysRevLett.133.036001}.

Of course, the microscopic description of these two materials is much more complicated than our simple model\textemdash for instance,  Cu intercalation also adds charge carriers in the case of $1T$-TiSe$_2$. Nevertheless, it is an interesting possibility that intercalation in nonmetallic CDW TMDs can provide a mechanism for a spatially inhomogeneous electron-phonon coupling.

\emph{Acknowledgements.}~
J.M and T.M were supported by the NSFC (12474218) and the Beijing Natural Science
Foundation (No. 1242022), and the work of R.S. was supported by Grant DE-SC0014671 funded by
the U.S. Department of Energy, Office of Science. Y.Z. and R.M.F. were supported by the Air Force Office of Scientific Research under Award No. FA9550-21-1-0423.

\bibliography{bib_diluteHols}


\newpage
\onecolumngrid
\clearpage
\setcounter{equation}{0}
\setcounter{figure}{0}
\renewcommand{\theequation}{S\arabic{equation}}
\renewcommand{\thefigure}{S\arabic{figure}}
\renewcommand{\thesubsection}{S\arabic{subsection}}

\def\avg#1{\langle#1\rangle}
\def\Re{\rm{Re}}
\def\Im{\rm{Im}}
\def\be{\begin{equation}} \def\ee{\end{equation}}
\def\bea{\begin{eqnarray}} \def\eea{\end{eqnarray}}
\def\PRB{Phys. Rev. B}
\def\PRA{Phys. Rev. A}
\def\PRL{Phys. Rev. Lett.}
\def\nn{\nonumber}
\def\pp{\parallel}

\begin{center}
{\bf{Supplemental Material for ``Supersolid Phase in the Diluted Holstein Model"}}
\end{center}
\begin{center}
{Jingyao Meng$^{1}$, Yuxi Zhang$^{2}$, Rafael M. Fernandes$^{2}$, Tianxing Ma$^{1,3,*}$ and R. T. Scalettar$^{4}$}
\end{center}
\begin{center}
\it{
$^{1}$School of Physics and Astronomy, Beijing Normal University, Beijing 100875, China

$^{2}$School of Physics and Astronomy, University of Minnesota, Minneapolis, MN 55455, USA

$^{3}$Key Laboratory of Multiscale Spin Physics (Ministry of Education), Beijing Normal University, Beijing 100875, China

$^{4}$Department of Physics and Astronomy, University of California,Davis, CA 95616,USA}

\end{center}

In this supplemental material we provide further discussion of disorder averaging, the spectral function $A(\omega)$ and the compressibility $\kappa$. To eliminate the possibility that the coexistence of
charge density wave and superconductivity is not a supersolid phase but instead results from averaging over realizations which are purely superconductivity and others which are purely charge density wave, we examine
the statistics of the individual realizations. 
A complementary way to examine the charge density wave gap is via
the momentum integrated single-particle spectral function, $A(\omega)$, which
can be obtained from the imaginary time dependent Greens function
computed in DQMC via the analytic continuation.  The compressibility $\kappa$ obtained through $\partial n/\partial\mu$ is also discussed.

\date{Version 16.0 -- \today}

\maketitle
 
\begin{center}
\textbf{1. Further Discussion of Disorder Averaging}
\end{center}

\begin{figure}[htbp]
\includegraphics[height=4.00in,width=4.00in,angle=0]{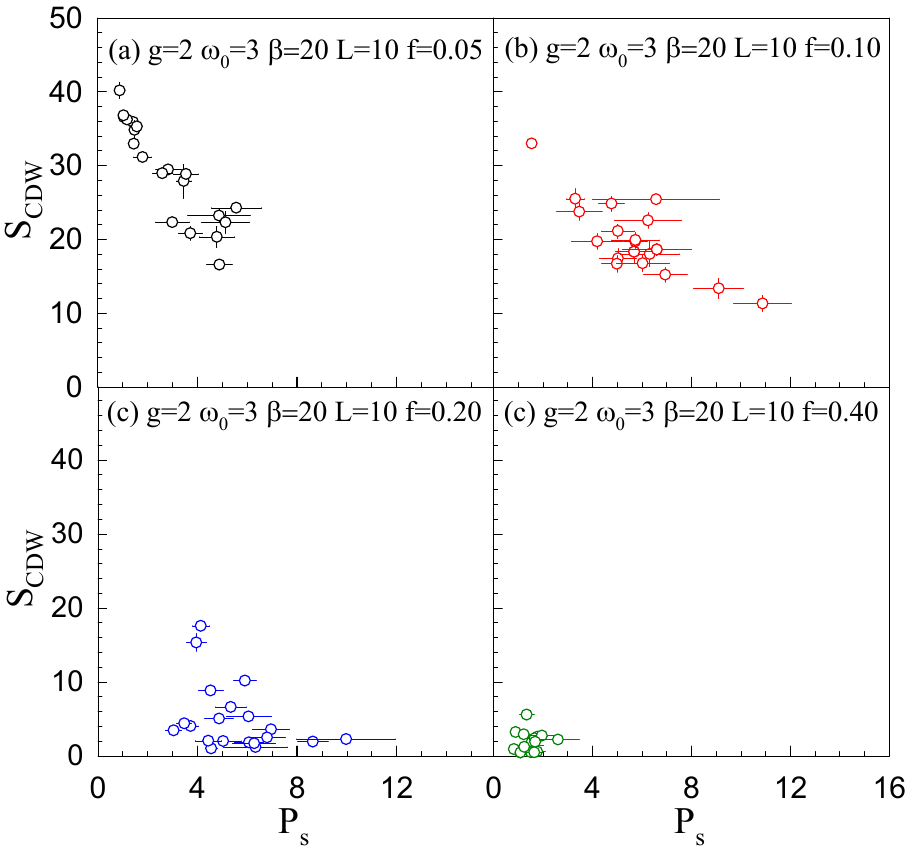}
\centering
\caption{Scatter plot of $S_{\rm CDW}$ and $P_{s}$ for $g=2$, $\omega_0=3$, $\beta=20$ with (a) $f=0.05$ in CDW phase (b) $f=0.10$ in SS phase and (c) $f=0.2$ in SC phase.
There is no evidence for a bimodal distribution of the order parameters, as would be the case if the simulation were falling
into distinct SC and CDW phases depending on the initial conditions.
The lattice size is $N=10\times 10$.
(d)  The metallic phase is characterized by small values of both order parameters.
}
\label{SMPsCDW}
\end{figure}

In the results reported in the main text, data were averaged over 10-20 disorder realizations. To eliminate the possibility that the coexistence of
CDW and SC is not a supersolid phase but instead results from averaging over realizations which are purely SC and others which are purely CDW, we examine
the statistics of the individual realizations.
We show a scatter plot of $S_{\rm cdw}$ and $P_{s}$ in Fig.~\ref{SMPsCDW} for 20 disorder realizations.
The supersolid phase is in panel (b), $f=0.10$.  We see no evidence for two distinct clumps of data,
one at large $S_{\rm cdw}$--small $P_s$ and the other
large $P_s$--small $S_{\rm cdw}$,
as would be expected if different realizations were individually in one phase or another.

\begin{figure}[htbp]
\includegraphics[height=2.50in,width=2.50in,angle=0]{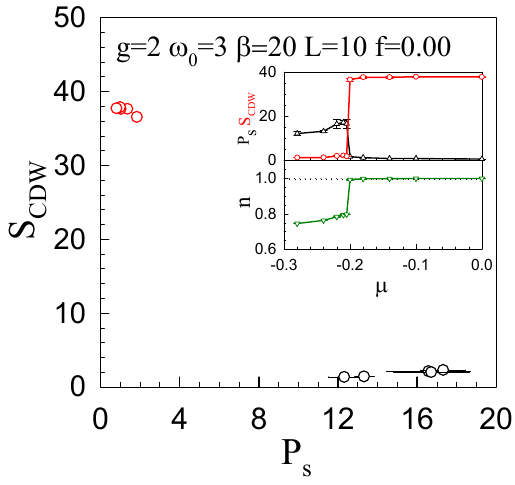}
\centering
\caption{Scatter plot of $S_{\rm CDW}$ and $P_{s}$ for $g=2$, $\omega_0=3$, $\beta=20$ with $\Delta \mu\in$[-0.28, 0.00]. The range of the CDW gap, and behaviors of $n$, $P_{s}$ and $S_{CDW}$ are shown in the inserts. The phase is either CDW ($\Delta \mu > -0.20$) or SC ($\Delta \mu < -0.20$).
The associated order parameters $S_{\rm cdw}$ and $P_s$ are never simultaneously large--  the scatter plot consists of
two distinct clumps associated with the CDW and SC phases.
}
\label{SMIQR}
\end{figure}

Figure \ref{SMIQR} provides additional perspective by showing a related scatter plot in the undiluted case, at $f=0.00$.
Here we do not do different realizations at the same parameters, but rather accumulate data as the chemical potential shift
from half-filling $\Delta \mu$ is varied. At $f=0.00$ the system is uniquely either a CDW (when $\Delta \mu$ is within
the CDW gap) or a SC (when  the system is sufficiently doped).  The scatter plot thus obtained shows two distinct clusters
of points, emphasizing the distinction with the SS in  Fig.~\ref{SMPsCDW}(b).

\begin{figure}[htbp]
\includegraphics[height=2.50in,width=5.00in,angle=0]{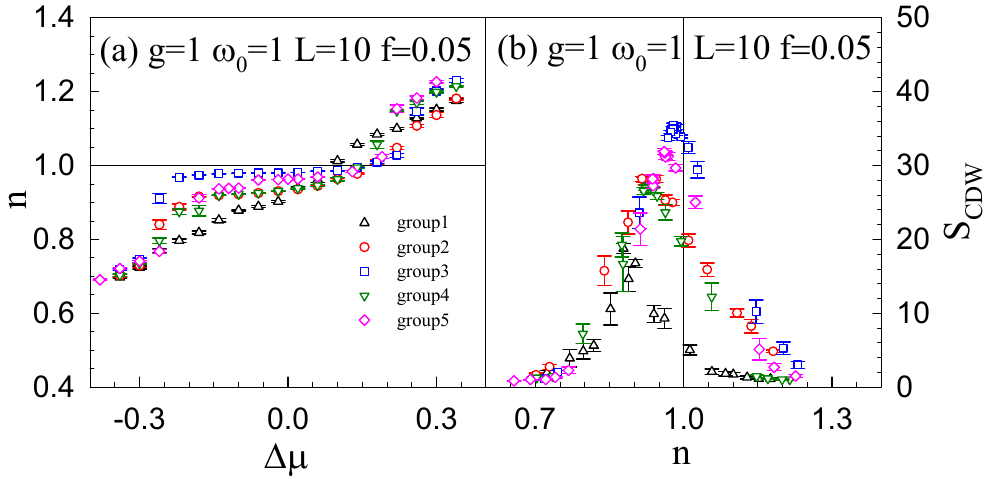}
\centering
\caption{(a) Density $\langle n \rangle$ as the function of shift in chemical potential $\Delta\mu$
away from $\mu_0$ for five groups of disorder realizations.  A plateau appears 
in each case.  (b) $S_{\rm cdw}$ as a function of $\langle n \rangle$ for the same
binning. The maximal value of the charge structure factor occurs at $\langle n \rangle <1$.
This reflects the fact that, at $f=0.05$, the phonon mode is absent on some of the sites.
}
\label{SMgap}
\end{figure}

In showing the vanishing of the compressibility in the CDW gap in Fig. {\color{red}5}.
of the main text, we did not disorder average, because the $\kappa=0$ region appears at a different $\Delta \mu$
for each realization,  somewhat obscuring the plateau structure.
Here in Fig.~\ref{SMgap} we provide further data by dividing the 20 realizations into five groups, each averaged over four realizations.
The plateau structure is still evident, as is the shift in the peak of
$S_{\rm cdw}$ away from $\langle n \rangle=1$.  If we were to simulate larger lattices (which is challenging since the
computational cost scales as $N^3 = L^6$) the data would self-average and all realizations/groups would exhibit a common
plateau at low $T$ (and hence far from $T_c$).  However, it should be noted that in the high precision studies possible on large lattices in
classical simulations of diluted Ising models, subtle non-self-averaging effects can be observed
in the critical exponents when very  close to the critical point \cite{berche2004bond,PhysRevB.58.2740,PhysRevE.58.2938,PhysRevLett.81.252}.

\begin{center}
\textbf{2. The Spectral Function}
\end{center}

Figure {\color{red}5} 
of the main text provided one insight into the CDW gap by examining
the vanishing if the compressibility $\kappa = \partial \langle n \rangle/\partial \mu$,
i.e.~a plateau in $\langle n \rangle$ vs.~$\mu$.
as will Fig.~\ref{SMkappa} below.
A complementary way to examine this is via
the momentum integrated single-particle spectral function, $A(\omega)$, which
can be obtained from the imaginary time dependent Greens function
computed in DQMC via the analytic continuation of
\begin{align}
\frac{1}{N} \sum_{\bf p} G({\bf p}=0,\tau)
= G({\bf r}=0,\tau)
= \int d\omega \, \frac{e^{-\omega \tau}}{e^{\beta \omega}+1} \, A(\omega)
\end{align}

Here, a gap in $A(\omega)$ can occur either from the formation of CDW or SC phases,
and an effective gap emerges when both types of order are present, a SS.  Its precise structure depends
on which orders are intertwined.
In the case where the SC gap is
isotropic $s$-wave and the competing order
is CDW, a straightforward diagonalization of the mean-field Hamiltonian showns that the effective gap is the quadrature sum of the SC and CDW gaps,
$\Delta_{\rm eff} \sim \sqrt{\Delta_{\rm sc}^2 + \Delta_{\rm cdw}^2}$ \cite{Machida1984,PhysRevB.82.014521}.

\begin{figure}[htbp]
\includegraphics[scale=0.32]{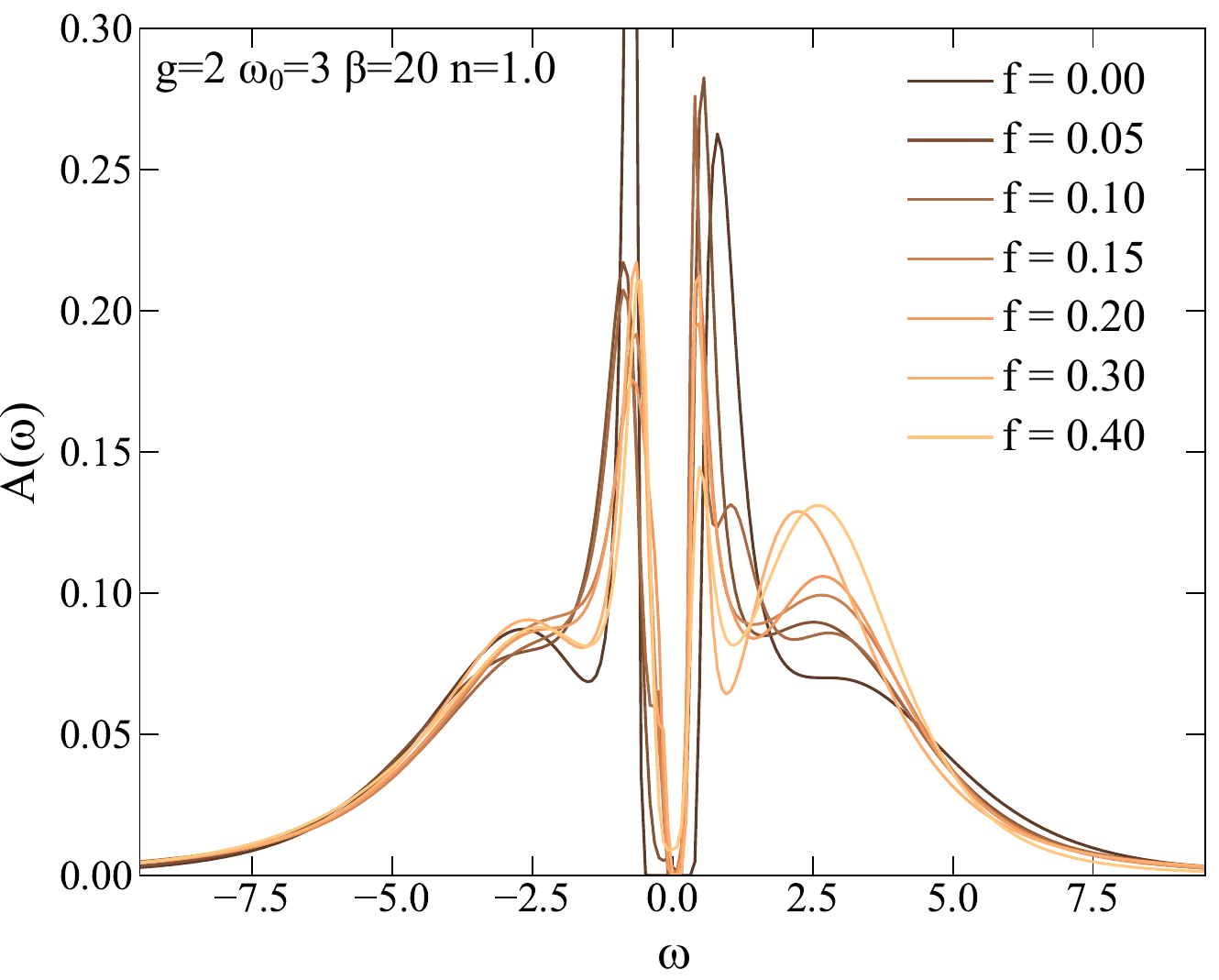}
\centering
\caption{Spectral function $A(\omega)$ for various $f$ at $g=2$, $\omega=3$, $\beta=20$. Curves under $f=$0.0 to 0.3 show clear gaps at $\omega = 0$, while strong dilution $f=0.4$ both destroy gap and long range orders.}
\label{SMAw}
\end{figure}

We calculate $A(\omega)$ through an analytic continuation of
DQMC imaginary time data for the single particle Green's function
$G(r=0,\tau)$ using ana-cont (\cite{kaufmann23}) and find a non-zero
gap is present for $g=2, \omega_0=3, \beta=20$ for a range of dilutions $0.0 < f
\lesssim 0.35$, and vanishes at $f \sim 0.40$  (See Fig.~\ref{SMAw}.)
This is consistent with the effective gap $\Delta_{\rm eff}$
discussion above, and provides further confirmation of the
supersolid phase:  We do not see $\Delta_{\rm eff}$ becoming small at an
intermediate dilution $f$,
as would occur if $\Delta_{\rm sc}$ and $\Delta_{\rm cdw}$
both vanished at a CDW-SC transition.  This suggests that there
a region of overlap where both order parameters are finite.

Although $A(\omega)$ is therefore not a good marker of the
evolution from CDW to SS to SC, the compressibility
$\kappa = \frac{d \langle n \rangle}{d \mu}$ does change character,
vanishing in the insulating CDW phase, but taking on a non-zero
value in the compressible SC region.  We find $\kappa$
becomes non-zero at dilution $f \sim 0.15$.  (See Fig.~\ref{SMkappa}.)
This is consistent with the phase diagram of Fig. {\color{red}4}(b). 

Besides, the maximum entropy (ME) method is a standard approach to obtain $A(\omega)$ on the real-frequency axis in strongly correlated system. Although the regularization based on Bayesian arguments may be unreliable at divergent fluctuations at continuum limit and $A(\omega)$ are sensitive to the fine structures of Green’s function at high $\omega$, our study focuses on the behavior of gap in low frequency region, avoiding the flaw in investigating high frequency information or sharp peaks \cite{PhysRevLett.126.056402,PhysRevB.101.085111,HUANG2023108863}.

\begin{center}
\textbf{3. Compressibility}
\end{center}

Finally, we show in Fig.~\ref{SMkappa}(a) the compressibility $\kappa$ obtained through $\partial n/\partial\mu$, the finite difference of density in two adjacent runs at different potential.
In the thermodynamic limit and at $T=0$, $\kappa$ vanishes in a gapped CDW phase \cite{PhysRevLett.130.226001,PhysRevA.74.013608}.
The $f=0.00$ and $f=0.05$ curves of Fig.~\ref{SMkappa} have $\kappa=0$.
In  the phase diagram of Fig. {\color{red}4}(b), 
the CDW phase terminates at $f=0.15$, and indeed
$\kappa$ becomes non-zero there. 
The data for $S_{\rm cdw}$ and its scaling in the main text demonstrate that the CDW is
still present at $f=0.10$, yet we see a small residual $\kappa$ there.
However, it is known that
on finite lattices and at nonzero temperatures, temperature broadening and finite size effects yield precisely such small non-zero
$\kappa$ close to the termination of the CDW phase \cite{PhysRevB.76.144513}.

Another key observation here is that at dilutions such as $f=0.20$ the system is
compressible (nonzero $\kappa$ in
Fig.~\ref{SMkappa}) but there is still a (SC) gap
as seen in  Fig.~\ref{SMAw}. 
The unusual phenomenon whereby $A(\omega=0)=0$ while $\kappa\neq0$ precisely indicates SC behavior-- the density of the system responds
to increase in the chemical potential
by adding {\it pairs} of fermions, but the {\it single particle} spectral function vanishes.

\begin{figure}[htbp]
\includegraphics[scale=0.32]{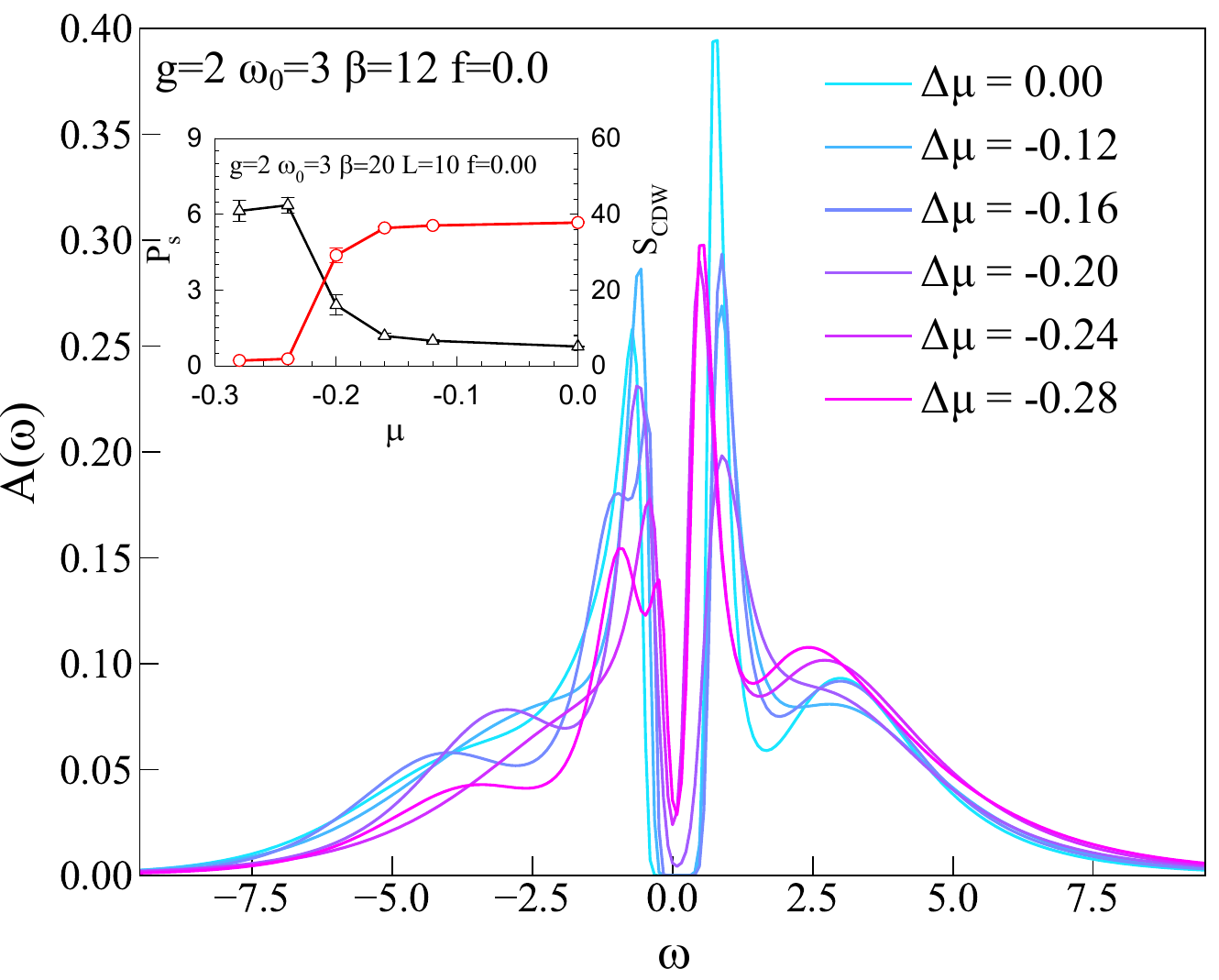}
\centering
\caption{(a) $A(\omega)$ for various potential at $\beta=12$. Gap vanishes around $\mu=-0.18$ under pretty high temperature. (b) $P_{s}$ and $S_{CDW}$ as the function of potential. CDW order has disappeared at $\mu=-0.20$.
}
\label{SMgap_mu}
\end{figure}

\begin{figure}[htbp]
\includegraphics[height=2in,width=4.2in,angle=0]{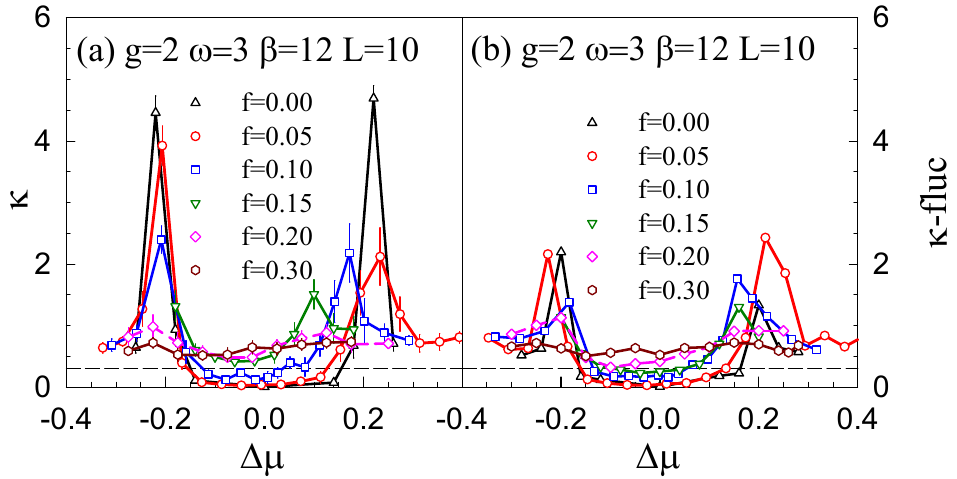}
\centering
\caption{(a) Compressibility $\kappa$ as the function of $\Delta\mu$ at $g=2$, $\omega=3$, $\beta=12$ and $L=10$. As $f$ increases, the $\kappa$ curve gradually flattens out and moves away from $f-$axis. At $f=0$, $\kappa$ is around 0.4 with $\Delta\mu$ arriving -0.18.
(b) Compressibility $\kappa$ obtained from summing the density-density correlations.  The two methods give
consistent results for the locations of the peaks, which signal the range
of the insulating plateau.  The quantitative values agree over most of the range,
except right at the peak.  Here the fluctuation method underestimates $\kappa$ due to
mesh size of our data, which never lands precisely on the critical $\Delta\mu$.
The finite difference approach is better able to capture the sharp peak since a large
change in density can occur even for chemical potentials which bracket the transition.}
\label{SMkappa}
\end{figure}

\end{document}